\documentclass[aps,twocolumn,showpacs,fleqn]{revtex4}

\usepackage{graphicx}
\usepackage{amsmath}
\usepackage{amsfonts}
\usepackage{amssymb}
\usepackage{epstopdf}

\newcommand{\mean}[1]{\left\langle #1 \right\rangle}
\def\ddt{\partial_{t}}

\renewcommand{\Im}{\text{Im}\,}
\renewcommand{\Re}{\text{Re}\,}
\renewcommand{\imath}{\mathrm{i}}

\newcommand{\PIs}{\Pi'}
\newcommand{\PIss}{\Pi''}
\newcommand{\PF}{\Phi}
\newcommand{\PFHI}{\tilde{\Phi}}

\newcommand\bop{\hat{b}}\newcommand\Bop{\hat{B}}\newcommand\cop{\hat{c}}\newcommand\nop{\hat{n}}

\newcommand\Eaks{\Delta_{\alpha k s}}
\newcommand\fermi{f}
\newcommand\tpulse{t_{\rm p}}

\begin{document}
\title{Non-adiabatic Electron Pumping through Interacting Quantum Dots}
\author{Alexander Croy}
\email{alexander.croy@chalmers.se}
\altaffiliation{Current address: Department of Applied Physics,
Chalmers University of Technology, S 41296 G\"{o}teborg, Sweden}

\author{Ulf Saalmann}
\affiliation{Max-Planck-Institute for the Physics of Complex Systems,
   N\"{o}thnitzer Str.~38, 01187 Dresden, Germany}

\author{Alexis R. Hern\'andez}
\affiliation{Instituto de F\'{\i}sica, Universidade Federal do Rio de Janeiro, 21941-970 Rio de Janeiro, Brazil}
	 
\author{Caio H. Lewenkopf}
\affiliation{Instituto de F\'{i}sica, Universidade Federal Fluminense, 24210-346 Niter\'{o}i, Brazil}
	
\date{\today}

\begin{abstract}\noindent
We study non-adiabatic charge pumping through single-level quantum dots taking into account Coulomb interactions. We show how a truncated set of equations of motion can be propagated in time by means of an auxiliary-mode expansion. This formalism is capable of treating the time-dependent electronic transport for arbitrary driving parameters. We verify that the proposed method describes very precisely the well-known limit of adiabatic pumping through quantum dots without Coulomb interactions. As an example we discuss pumping driven by short voltage pulses for various interaction strengths. Such finite pulses are particular suited to investigate transient non-adiabatic effects, which may be also important for periodic drivings, where they are much more difficult to reveal. 
\end{abstract}
	
\pacs{73.63.Kv, 73.23.Hk, 72.10.Bg}

   \maketitle
\section{Introduction}\label{sec:Intro}

In 1983 Thouless \cite{th83} proposed a simple pumping mechanism to produce, even in the absence of an external bias, a  quantized electron current through a quantum conductor by an appropriate time-dependent variation of the system parameters. 
Experimental realizations of quantum pumps using quantum dots (QDs) were already reported in the early 90's \cite{pola+92,mana+94}. More recently, due to the technological advances in nano-lithography and control, such experiments have risen to a much higher sophistication level, making it possible to pump electron \cite{swma+99,dima+03,lebu+05} and spin \cite{wapo+03} currents through open nanoscale conductors, as well as through single and double QDs \cite{shso+06,fufa+07,buka+08,funi+08}.

Early theoretical investigations where devoted to the adiabatic pumping regime within the single-particle approximation \cite{br98,zhsp+99,enah+02}. This is well justified for experiments with open QDs, where interaction effects are believed to be weak \cite{albr+02} and the typical pumping parameters are slow with respect the characteristic transport time-scales, such as the electron dwell time $\tau_{\rm d}$. 
This time-scale separation enormously simplifies the analysis of the two-time evolution of the system.  
Within the adiabatic regime,  inelastic and dissipation \cite{mobu01} effects of currents generated by quantum pumps were analyzed. 
Furthermore, issues like counting statistics \cite{mami01}, memory effects \cite{roci10}, and generalizations of charge pumping to adiabatic quantum spin pumps were also proposed and studied \cite{shch01,much+02,male+04}.

Non-adiabatic pumping has been theoretically investigated within the single-particle picture, either by using Keldysh non-equilibrium Green's functions (NEGF) with an optimal parametrization of the carrier operators inspired by bosonization studies \cite{vaam+01}, or by a Flouquet analysis of the $S$-matrix obtained from the scattering approach \cite{mobu02}. While the first approach renders complicated integro-differential equations for the Green's functions associated to the transport, the second one gives a set of coupled equations for the Flouquet operator. It is worth to stress that, in both cases the single-particle picture is crucial to make the solution possible and it is well established that both methods are equivalent \cite{armo06,mule07}.

Several works have provided a quite satisfactory description of quantum pumping for weakly interacting systems. In contrast, the picture is not so clear for situations where interaction effects are important. Different approximation schemes have been proposed to deal with pumping in the presence of interactions and to address charging effects, which are not accounted for in a mean-field approximation.
Typically, two limiting regimes have been studied, namely, the one of small pumping frequencies $\Omega$, such that $\Omega\tau_{\rm d}\ll 1$ (adiabatic limit) \cite{cian+03,ao04,spgo+05,seor06,fisi08,arye+08,hepi+09} and the one of very high frequencies, $\Omega \tau_{\rm d} \gg 1$ (sudden or diabatic limit) \cite{hawe+01,saco+06,brbu08}. Nonadiabatic pumping is mainly studied as a side effect of photon-assisted tunneling \cite{blha+95,ooko+97,to05}, where $\Omega \tau_{\rm d} \gg 1$.
 
Unfortunately, it is quite cumbersome to calculate corrections to these limit cases. For instance, the analysis of higher-order corrections to the adiabatic approximation for the current gives neither simple nor insightful expressions \cite{hepi+09}. In addition to the theoretical interest, a comprehensive approach bridging the limits of $\Omega \tau_{\rm d} \gg 1$ and $\Omega \tau_{\rm d} \ll 1$ has also a strong experimental motivation: Most current experimental realizations of quantum pumping deal with QDs in the Coulomb blockade regime and $\Omega\tau_{\rm d} \sim 1$. This regime was recently approached (from below)  by means of a diagrammatic real-time transport theory with a summation to all orders in $\Omega$ \cite{cago+09}. However, the derivation implied the weak tunnel coupling limit, whereas experiments  
\cite{blka+07,wrbl+08,kaka+08a,kaka+08,wrbl+09} typically rely on tunnel coupling variations which include both weak and strong coupling.  

To address the above mentioned issues and to account for the different time scales involved it is natural to use a propagation method in the \emph{time domain} \cite{zhma+05,kust+05,wesc+06,mogu+07,myst+09}. In this work we express the current operator in terms of density matrices in the Heisenberg representation. We obtain the pumped current by truncating the resulting equations-of-motion for the many-body problem. The time-dependence is treated exactly by means of an auxiliary-mode expansion \cite{wesc+06,crsa09a}. This approach provides a quite amenable path to circumvent the usual difficulties of dealing with two-time Green's functions \cite{crsa09a}. Moreover, it has been successfully applied to systems coupled to bosonic reservoirs \cite{meta99} and to the description of time-dependent electron-transport using generalized quantum master equations for the reduced density matrix \cite{wesc+06,jizh+08,crsa11}. Since the auxiliary-mode expansion is well controlled \cite{crsa09crsa10}, the accuracy of our method is determined solely by the level of approximation used to treat the many-body problem.
 
The formalism we put forward is illustrated by the study of the charge pumped through a QD in the Coulomb-blockade regime by varying its resonance energy and couplings to the leads. The external drive is parametrized by a single pulse, whose duration and amplitude can be arbitrarily varied. By doing so, the formalism is capable to reproduce all known results of the adiabatic limit and to explore transient effects beyond this simple limit.
 
The paper is organized as follows. In Sec.\ \ref{sec:model} we present the resonant-level model, as well the theoretical framework employed in our analysis. In Sec.\ \ref{sec:Prop} we introduce the general propagation scheme, suitable to calculate the pumping current at the adiabatic regime and beyond it. Next, in Sec.\ \ref{sec:App}, we discuss few applications of the method. Finally, in Sec.\ \ref{sec:conclusion} we present our conclusions.

\section{Time-dependent Interacting Resonant-Level Model}
\label{sec:model}
   
The standard model to address electron transport through QDs is the Anderson interacting single-resonance model coupled to two reservoirs, one acting as a source and the other as a drain. Despite its simplicity, the model provides a good description for Coulomb-blockade QDs and for QDs at the Kondo regime, where the electrons are strongly correlated. In this paper we address the Coulomb-blockade regime, for QDs whose typical line width $\Gamma$ is much smaller than the QD mean level spacing $\delta$, justifying the use of the Anderson single-resonance model. In addition, in the Coulomb blockade regime $\Gamma$ is much smaller than the resonance charging energy $U$.

\subsection{Setup}
\label{sec:setup}

The total Hamiltonian is given by the usual threefold decomposition into a quantum dot Hamiltonian $H_{\rm dot}$, a Hamiltonian $H_{\rm leads}$ representing the leads, and a coupling term $H_{\rm coup}$, namely
   \begin{subequations}\label{eq:Hamiltonian}
   \begin{equation}
     H = H_\mathrm{dot} + H_\mathrm{leads} + H_\mathrm{coup}\,.
     \label{eq:totalHam}
   \end{equation}
The QD is modeled by a single level of energy $\varepsilon_{\rm d}(t)$, which can be occupied by spin-up and spin-down electrons, which interact through a contact interaction of strength $U$. The QD Hamiltonian reads
   \begin{align}\label{eq:NEGFCBHam}
     H_{\rm dot} 
     &=\sum\limits_{s=\uparrow,\downarrow} \varepsilon_{\rm d}(t) \hat{n}_s +
     U \hat{n}_{\uparrow} \hat{n}_{\downarrow} \;,
   \end{align}
where $\hat{n}_{s}=\cop^\dagger_{s}\cop_{s}$, $\cop^\dagger_{s}$ and $\cop_{s}$ stand for electron number, creation and annihilation operator for the respective spin state $s=\uparrow,\downarrow$ in the dot.

The two reservoirs, labeled as $L$ (left) and $R$ (right), are populated by non-interacting electrons, whose Hamiltonian reads
   \begin{equation}
     H_\mathrm{leads} = \sum_{\alpha \in \mathrm{L},\mathrm{R}} 
     \sum_{ks} \varepsilon_{\alpha k}(t) \bop^\dagger_{\alpha ks} \bop_{\alpha ks}^{}\;,
     \label{eq:ResHamilOp}
   \end{equation}
where $\{\bop^\dagger_{\alpha k s}\}$ and $\{\bop_{\alpha k s}\}$ stand for the electron creation and annihilation operators for the $\alpha$-reservoir state $ks$, respectively.   
The reservoir single-particle energies have the general form $\varepsilon_{\alpha k}(t) = \varepsilon^0_{\alpha k} + \Delta\varepsilon_{\alpha}(t)$ with the $\Delta\varepsilon_{\alpha}$ accounting for a time-dependent bias. The stationary current due to a time-dependent bias was already addressed several years ago \cite{jawi+94}. For pumping, we take $\Delta\varepsilon_{\alpha}(t)=0$, as usual.

Finally, the coupling Hamiltonian is given by
   \begin{eqnarray}
     H_\mathrm{coup} &=& \sum_{\alpha k}\sum_{s} 
     T_{\alpha k}(t) \,\bop^\dagger_{\alpha ks} \cop^{}_s
     + \rm{H.c.} \;,
     \label{eq:TunnHam}
   \end{eqnarray}
   \end{subequations}
with $\{T_{\alpha k}\}$ denoting the coupling matrix element between the QD and the reservoir $\alpha$.

\subsection{Equation of motion approach}
\label{sec:EOMTIntro}
We are interested in the electronic current from reservoir $\alpha$ to the QD state $s$, which can be obtained from current operator
\begin{align}\label{eq:DefCurrOp}
\hat{J}_{\alpha s}(t) \equiv {} &\imath \sum_{k} \left[
T_{\alpha k}(t) \bop^\dagger_{\alpha k s}(t) \cop_{s}(t)
\right.\notag\\ &\qquad\left.
-  T^*_{\alpha k}(t) \cop^\dagger_{s}(t) \bop_{\alpha k s}(t) \right]\,.
\end{align}
Here and in the following we use units where the elementary charge $e = 1$ and the reduced Planck constant 
$\hbar=1$, unless otherwise indicated.
To calculate $\hat{J}_{\alpha s}(t)$ we use the 
following equations of motion, which are obtained from the Hamiltonian 
[Eqs.\,\eqref{eq:Hamiltonian}] by means of the Heisenberg equation,
\begin{subequations}\label{eq:EOMTHeisenberg}
\begin{align}
  \imath \partial_t \bop_{\alpha k s}(t) 
  		= {} & \varepsilon_{\alpha k}(t) \bop_{\alpha k s}(t)
                                  + T_{\alpha k}(t) \cop_{s}(t)\;,
                                  \label{eq:EOMTHeisenberg1}\\
  \imath \partial_t \cop_{s}(t) 
  		= {} & \varepsilon_s(t) \cop_{s}(t) + U \cop_{s}(t) \nop_{\bar{s}}(t)\notag\\
       & + \sum\limits_{\alpha k} T^*_{\alpha k}(t) \bop_{\alpha k s}(t)\;,
       \label{eq:EOMTHeisenberg2}\\
  \imath \partial_t \nop_{\bar{s}}(t) 
  		= {} & \sum\limits_{\alpha k} \left[
    	- T_{\alpha k}(t) \bop^\dagger_{\alpha k \bar{s}}(t) \cop_{\bar{s}}(t) 
	 \right.\notag\\
    &\qquad\left.
    	+ T^*_{\alpha k}(t) \cop^\dagger_{\bar{s}}(t) \bop_{\alpha k \bar{s}}(t)
  \right]\;.\label{eq:EOMTHeisenberg3}
\end{align}
\end{subequations}
Analogous equations hold for $\bop^{\dagger}_{\alpha k s}$ and $\cop^{\dagger}_{s}$.

In the spirit of the scheme introduced by Caroli and co-workers \cite{caco+71}, we assume an initially uncorrelated density operator of the combined system, {\it i.\,e.}, we set $T_{\alpha k}(t_0)\to 0$ for $t_0\to -\infty$.
Further, we apply the so-called wide-band limit \cite{haja07}, where the square of the tunneling element $T_{\alpha k}$ is inversely proportional to the density of states $\rho_\alpha$ at energy $\varepsilon_{\alpha k}$.
By means of the lead Green function \cite{haja07}
\begin{equation}\label{eq:DefLeadGF}
	g_{\alpha k}(t,t') = 
		\exp\left[ -\imath \int^{t}_{t'} dt'' \varepsilon_{\alpha k}(t'')\right]
\end{equation}
we can define the decay rate
\begin{subequations}\label{eq:GammaWBL}
\begin{align}\label{eq:DefLevWidFun1}
	\Gamma_\alpha (t,t') \equiv {} & \sum_k T_{\alpha k} (t) g_{\alpha k}(t,t') T^*_{\alpha k}(t')\;,
\end{align}
which becomes local in time in the  wide-band limit, namely
\begin{align}\label{eq:DefLevWidFun2}
	\Gamma_\alpha (t,t') 
	& = \int\mathrm{d}\varepsilon_{k} \rho(\varepsilon_{k})
	T_{\alpha k} (t) g_{\alpha k}(t,t') T^*_{\alpha k}(t')\notag\\
	& = \Gamma_\alpha (t) \delta(t-t')\;.
\end{align}
\end{subequations}
In the following we replace the sum in Eq.\ \eqref{eq:DefLevWidFun1} by the expression involving the $\delta$-function in Eq.\ \eqref{eq:DefLevWidFun2}. 

The equation of motion for the reservoir operators $\bop_{\alpha k s}$, Eq.~\eqref{eq:EOMTHeisenberg1}, is now readily integrated, yielding
\begin{subequations}\label{eq:EOMTResOp}
\begin{equation}
	\bop_{\alpha k s}(t) =
	\Bop_{\alpha k s}(t)
	-\imath \int\limits^t_{t_0} dt' g_{\alpha k}(t,t') T_{\alpha k}(t') \cop_{s}(t')\;,
\end{equation}
where we have used the lead Green functions, Eq.\ \eqref{eq:DefLeadGF}, and introduced
\begin{equation}
	\Bop_{\alpha k s}(t) \equiv g_{\alpha k}(t,t_0) \bop_{\alpha k s}(t_0)\;. \label{eq:BDef}
\end{equation}
\end{subequations}
Equations \eqref{eq:EOMTResOp} are used to rewrite Eq.\ \eqref{eq:EOMTHeisenberg2} as
\begin{align}
  \imath  \partial_t \cop_{s}(t)
		= {} & \left[
			\varepsilon_s(t) + U \nop_{\bar{s}}(t) 
			- \imath \frac{\Gamma(t)}{2} \right] \cop_{s}(t) \notag\\
		&+ \sum\limits_{\alpha k} T^*_{\alpha k}(t) \Bop_{\alpha k s}(t)\,.
		\label{eq:cEOM}
\end{align}
Here the wide-band limit, Eq.\ \eqref{eq:GammaWBL}, is employed to obtain the decay term, proportional to $\Gamma(t)=\sum_{\alpha}\Gamma_{\alpha}(t)$.
Similarly, we can rewrite Eq.\ \eqref{eq:EOMTHeisenberg3} as
\begin{align}
  \imath  \partial_t \nop_{\bar{s}}(t) 
   	= {}  \sum\limits_{\alpha k} &\left[ 
    - T_{\alpha k}(t) \Bop^\dagger_{\alpha k \bar{s}}(t) \cop_{\bar{s}}(t) \right. \label{eq:NsbarEOM}\\
    &\left.
    \;+ T^*_{\alpha k}(t) c^\dagger_{\bar{s}}(t) \Bop_{\alpha k \bar{s}}(t)
  \right] 
  - \imath \Gamma(t) \nop_{\bar{s}}(t)\;.\notag
\end{align}
Here again the time integral of $\bop_{\alpha k s}(t)$ is reduced to a decay width due to the wide-band limit \eqref{eq:GammaWBL}.

\subsection{Expectation values and truncation schemes}\label{sec:ExpVal}

The expression for the time-dependent current is given by the expectation value of 
the current operator $\hat{J}_{\alpha s}$ defined in Eq.\ \eqref{eq:DefCurrOp}.
As will become clear later on, it is useful to write this expectation value as
\begin{equation}\label{eq:CurrPi}
	J_\alpha(t) = \frac{e}{\hbar}\,\sum_{s}\mean{\hat{J}_{\alpha s}(t)} = \frac{2 e}{\hbar}\, \Re \sum_{s}  \Pi_{\alpha s} (t)
\end{equation}
with the {\it current matrices of the first order}
\begin{equation}\label{eq:PiDef}
\Pi_{\alpha s}(t) \equiv  \imath\sum_{k} T_{\alpha k}(t) 
\mean{ \bop^\dagger_{\alpha k s}(t) \cop_{s}(t) }\;.
\end{equation}
These current matrices are an essential ingredient of our propagation scheme,
which is based on finding equations of motion for $\Pi_{\alpha s}$. Such equations have been derived starting from a NEGF formalism for non-interacting electrons \cite{crsa09a}.

Exactly as for the operator equations above we can use $\bop_{\alpha ks}$ from Eq.\ \eqref{eq:EOMTResOp} and employ the wide-band limit \eqref{eq:GammaWBL} for the current matrices defined in Eq.\ \eqref{eq:PiDef}.
This leads to the following decomposition
\begin{subequations}\label{eq:NEGFDefPi1}
   \begin{align}
	\Pi_{\alpha s} ( t ) &=  \PIs_{\alpha s}(t) + \sum_k T_{\alpha k}(t) \PIss_{\alpha k s}(t) \,,\\
	\PIs_{\alpha s}(t) &= -\frac{\Gamma_\alpha(t)}{2} \mean{ \nop_{s}(t) } \,,\\
	\PIss_{\alpha k s}(t) &= \imath \mean{ \Bop^\dagger_{\alpha k s}(t) \cop_{s}(t)}\,.
	\end{align}
\end{subequations}
Having derived all relevant equations of motion for the operators we can specify the respective equations for the two contributions $\PIs$ and $\PIss$.
The term $\PIs$ is the simplest and is basically given by the equation of motion for $\nop_{s}$, cf.\ Eq.\ \eqref{eq:NsbarEOM}.
The corresponding equation for the occupation $n_{s}(t)\equiv\mean{ \nop_s (t)}$ reads
\begin{equation}\label{eq:EOMTPopulation}
\partial_t n_s (t) = 2\Re \sum_{\alpha k} T_{\alpha k}(t) \PIss_{\alpha k s}(t) 
	- \Gamma(t) n_{s}(t).
\end{equation}
The above relation can be viewed as the charge conservation equation for the QD. The rate by which the charge in the QD changes is equal to the total electronic currents. The first term at the {\it r.h.s.}~of the equation can be interpreted as the current flowing into the QD, whereas the second term gives the current flowing out.

Since we do not consider a spin-dependent driving or spin-polarized initial states it is $n_{s}(t)=n_{\bar{s}}(t)$. This relation is not explicitly used in the derivation, but is employed as a consistency check throughout the analysis.

The evaluation of $\PIss$ requires the solutions for both, the lead operator $\bop_{\alpha k s}$  and the dot operator $\cop_{s}$. Using those, we  write
\begin{align}
\partial_t \PIss_{\alpha k s}(t)
	= {} & \imath \Eaks(t)\, \PIss_{\alpha k s}(t) \notag\\
	&+ T^*_{\alpha k}(t) f_{\alpha k}
	-\imath U \PF_{\alpha k s}(t)\;.	\label{eq:Pi2EOM}
\end{align}
Here we have introduced the abbreviation
\begin{equation}\label{eq:Deltaks}
\Eaks(t)\equiv\varepsilon^0_{\alpha k} 
	- \left[\varepsilon_s(t) 
	- \imath \frac{\Gamma(t)}{2}\right]
\end{equation}
and used that
\begin{align}
\mean{\Bop^{\dagger}_{\alpha k s}(t)\Bop_{\alpha k s}(t)} 
&=\mean{\bop^{\dagger}_{\alpha k s}(t_{0})\bop_{\alpha k s}(t_{0})}\notag\\
&= \fermi_{\alpha}(\varepsilon_{k})\equiv f_{\alpha k}
\end{align}
with $\fermi_{\alpha}(\varepsilon)$ the Fermi function describing the equilibrium occupation of lead $\alpha$. The last term in Eq.\ \eqref{eq:Pi2EOM} uses the {\it auxiliary current matrices of the second order}
\begin{equation}
\PF_{\alpha k s} \equiv \imath \mean{ \Bop^\dagger_{\alpha k s}(t) \cop_{s}(t) \nop_{\bar{s}}(t) }\;,
\end{equation}
which will be subject to further approximations in the following.

Before we turn to the approximations, we would like to briefly discuss the physical meaning of $\PF_{\alpha k s}$. The equation of motion for the two-electron density matrix $\mean{ \nop_s (t) \nop_{\bar{s}} (t)}$ reads 
\begin{align}
	\partial_t \mean{ \nop_s (t) \nop_{\bar{s}} (t)} 
	= {} & - 2 \Gamma(t) \mean{ \nop_s (t) \nop_{\bar{s}} (t)} 
	 \notag\\ &
	+ 2\Re \sum\limits_{\alpha k s} T_{\alpha k s}(t) \PF_{\alpha k s}(t)\;,
	\label{eq:2eDMEOM}
\end{align}
which follows from Eq.\ \eqref{eq:NsbarEOM}. The two-electron density matrix may be interpreted as the occupation of one quantum-dot level under the condition that the other one is occupied. The rate of change of this conditional occupation is consequently given by tunneling into and out of the respective dot state under the same condition. The latter process is described by the first term on the {\it r.h.s.} of Eq.\ \eqref{eq:2eDMEOM}. The former process is governed by the auxiliary current matrices $\PF_{\alpha k s}$, which can be rewritten in the suggestive form
\begin{equation}
2\Re \sum\limits_{k} T_{\alpha k s}(t) \PF_{\alpha k s}(t)
= \mean{ \hat{J}_{\alpha s}(t) \nop_{\bar{s}} (t)}\;.
\end{equation}
Consequently, the current matrices $\PF_{\alpha k s}$ describe the {\it conditional current} 
from reservoir $\alpha$ into the quantum-dot level with spin $s$.

\subsubsection{Hartree-Fock approximation}
\label{sec:HFapp}

The simplest approximation to $\PF_{\alpha k s}$ consists in using the following factorization
\begin{align}
\PF_{\alpha k s}^{\rm HF}(t) & \equiv \imath \mean{  \Bop^\dagger_{\alpha k s}(t) \cop_{s}(t)} \mean{ \nop_{\bar{s}}(t) }
  \notag\\
	& = n_{\bar{s}}(t)\; \PIss_{\alpha k s}(t)\;.
\end{align}
Inserting this expression into Eq.\ \eqref{eq:Pi2EOM}, results in the following
equation of motion
\begin{align} \label{eq:PiEOMHartree}
\partial_t \PIss_{\alpha k s}(t) 
	= {} & \imath \left[  \Eaks(t)	- U n_{\bar{s}}(t)
	\right] \PIss_{\alpha k s}(t) \notag\\ &
	+ T^*_{\alpha k}(t) f_{\alpha k}.
\end{align}
This result is equivalent to the Hartree-Fock approximation applied to the Anderson model standard two-electron Green function \cite{haja07}. As any mean field approach, it does not lead to a double resonance Green function, which is required to properly account for charging effects. Hence, as it is well known, a good description of the Coulomb-blockade regime requires going beyond this level of truncation in the equations of motion.

\subsubsection{Hubbard I approximation}
\label{sec:HIapp}

Instead of factorizing $\PF_{\alpha k s}$ directly, we proceed by deriving its equation of motion. By means of Eqs.\ \eqref{eq:EOMTHeisenberg} we get
\begin{widetext}
\begin{align}
\partial_t \PF_{\alpha k s}(t) = {} &
	\imath\left[ \Eaks - U -\imath\Gamma(t)\right] \PF_{\alpha k s}(t)
	 +\sum\limits_{\alpha',k'} T^*_{\alpha' k'}(t) 
		\mean{ \Bop^\dagger_{\alpha k s}(t) \Bop_{\alpha' k' s}(t) \nop_{\bar{s}}(t)}\notag\\
	&
	 + \sum\limits_{\alpha',k'} \left[
		T_{\alpha' k'}(t)
		\mean{ \Bop^\dagger_{\alpha k s}(t) \cop_{s}(t) \Bop^\dagger_{\alpha' k' \bar{s}}(t) \cop_{\bar{s}}(t) }
		- T^*_{\alpha' k'}(t)
		\mean{ \Bop^\dagger_{\alpha k s}(t) \cop_{s}(t) c^\dagger_{\bar{s}}(t) \Bop_{\alpha' k' \bar{s}}(t) } \right].
	\label{eq:PhiEOM}
\end{align}
\end{widetext}
Note that the term proportional to $U$ has only four operators in the expectation values because of $\nop_{\bar{s}}\nop_{\bar{s}}=\nop_{\bar{s}}$.

The approximation consists in neglecting matrix elements involving opposite spins, which renders the following factorizations
\begin{align}
\mean{ b^\dagger_{\alpha k s}(t_0) \bop_{\alpha' k' s}(t_0) \nop_{\bar{s}}(t)}
			&\approx 
			f_{\alpha k} \delta_{\alpha\alpha'}\delta_{kk'} n_{\bar{s}}(t)\,,\notag\\
\mean{ \bop^\dagger_{\alpha k s}(t_0) \cop_{s}(t) 
	    	 \bop^\dagger_{\alpha' k' \bar{s}}(t_0) \cop_{\bar{s}}(t) }  &\approx 0\,,\notag\\
\mean{ \bop^\dagger_{\alpha k s}(t_0) \cop_{s}(t) 
	     \cop^\dagger_{\bar{s}}(t) \bop_{\alpha' k' \bar{s}}(t_0) }  &\approx 0\,.
\end{align}
This approximation for the density matrices is equivalent to the truncation scheme employed in the NEGF approach used for the study of Coulomb blockade regime (high-temperature limit of the Anderson model) \cite{haja07}.

As a result of the factorization, we obtain the following compact equation of motion for the approximated second-order current matrices
\begin{align}
\partial_t \PFHI_{\alpha k s}(t) = {}&
	\imath\left[ \Eaks - U - \imath\Gamma(t)\right] \PFHI_{\alpha k s}(t) \notag\\
	&+ T^*_{\alpha k}(t)\, f_{\alpha k}\, n_{\bar{s}}(t)\;.
	\label{eq:PiEOMHubbard}
\end{align}
The equations of motion for $n_{s}(t)$ [Eq.\ \eqref{eq:EOMTPopulation}],
$\PIss_{\alpha k s}(t)$ [Eq.\ \eqref{eq:Pi2EOM} with $\PF_{\alpha k s}$ replaced by $\PFHI_{\alpha k s}$] and for $\PFHI_{\alpha k s}(t)$ [Eq.\ \eqref{eq:PiEOMHubbard}] form a closed set of equations, which can be solved by means of an auxiliary-mode expansion discussed below.

\section{Auxiliary Mode Propagation Scheme}
\label{sec:Prop}
The general idea of the auxiliary-mode expansion consists in making use of a contour integration and the residue theorem to perform the energy integration, for instance, in Eq.\ \eqref{eq:NEGFDefPi1}. To this end the Fermi function is expanded in a sum over simple poles (or auxiliary modes) and the respective integrals are given as finite sums, cf.\ Appendix \ref{sec:AppExp}.

The transition to auxiliary modes (denoted by the index $p$) is facilitated by the following set of rules
\begin{subequations}\label{eq:rules}
\begin{align}
	\varepsilon_{\alpha k}(t) 
		&\longrightarrow \chi^+_{\alpha p}(t)\;, \\
	T^*_{\alpha k}(t) f_{\alpha k}
		&\longrightarrow T_{\alpha}(t) \left(\frac{-\imath}{\beta}\right) \;,\\
	\sum_k T_{\alpha k}(t) \PIss_{\alpha k s}(t)
		&\longrightarrow \frac{1}{4}\Gamma_{\alpha}(t)
		 + T_{\alpha}(t) \sum_p \PIss_{\alpha s p} (t)\;,
\end{align}
\end{subequations}
which are derived in Appendix \ref{sec:AppExp}. The first rule replaces the
reservoir energy $\varepsilon_{\alpha k}$ by the (complex) pole $\chi^+_{\alpha p}$
of the expansion, cf.\ Eq.\ \eqref{eq:auxmodesdef}. The second rule replaces the Fermi function by the respective weight, which is the same for all auxiliary modes. Finally, the third rule provides the actual expansion for the current matrices.

Applying these rules, the current matrices become
\begin{subequations}
\begin{align}
	\Pi_{\alpha s} &= \PIs_{\alpha s}(t) + \PIss_{\alpha s}(t) \,,\\
	\PIs_{\alpha s}(t) &= -\frac{\Gamma_\alpha(t)}{2} n_{s}(t) \,, \\
	\PIss_{\alpha s} (t) &=  \frac{\Gamma_{\alpha}(t)}{4} 
			+ T_{\alpha}(t)\sum\limits_{p} \PIss_{\alpha s p} (t)\,. \label{eq:PIss}
\end{align}
\end{subequations}
The equation of motion for the auxiliary matrix $\PIss_{\alpha s p}$
is obtained from Eq.\ \eqref{eq:Pi2EOM}. One arrives at  
\begin{eqnarray}
	\imath \partial_{t} \PIss_{\alpha s p} (t) 
       &=& 
       \left[ \varepsilon_s(t) -\imath \Gamma(t)/2 - \chi^+_{\alpha,p}(t) \right]
       	\PIss_{\alpha s p} (t) \nonumber\\
    && +\frac{1}{\beta}T_{\alpha}(t) + U \:\PF_{\alpha s p} (t)	\,.
    \label{eq:AuxEOMTPi}
\end{eqnarray}
The equations of motion for the auxiliary matrices $\PF_{\alpha s p}$ are quite similar to those of Eq.\ \eqref{eq:PIss}, namely,
\begin{align}
	\imath \ddt\PFHI_{\alpha s p} (t) 
       = {} &  \left[ \varepsilon_s(t) + U - \chi^+_{\alpha,p}(t)  - \imath \frac{3 \Gamma(t)}{2}\right] \PFHI_{\alpha s p} (t)
       	\nonumber\\
      &+ \frac{1}{\beta}T_{\alpha}(t)\, n_{\bar{s}}(t)\;. \label{eq:AuxEOMTPhi}
\end{align}
The solution of the above equations still requires a complete description 
of the population dynamics given by $n_{s}(t)$. The latter can be directly
obtained from Eq.\ \eqref{eq:EOMTPopulation} in terms of the current matrices
\begin{align}\label{eq:AuxEOMTPopulation}
   \ddt n_{s}(t) &= -\Gamma(t)\, n_{s}(t) + 2\Re\sum_{\alpha}\PIss_{\alpha s}(t) \;.
\end{align}   
This concludes the derivation of the auxiliary mode propagation scheme. The set of equations \eqref{eq:AuxEOMTPi} to \eqref{eq:AuxEOMTPopulation}, with initial conditions $n_s(t_0)=0$, $\PIss_{\alpha s p} (t_0)=0$, and $\PFHI_{\alpha s p} (t_0) =0$, can be solved numerically using standard algorithms.
Before the desired time dependence of the parameters $\varepsilon_s(t)$ and
$\Gamma(t)$ sets in, the system has to be propagated until a steady state is reached. 
In this way, transient effects arising from the choice of the initial state are avoided.
For convenience we derive in Appendix \ref{sec:AppStat} the expressions for the stationary occupations, which may also be used as initial values for $n_s$.

\section{Non-adiabatic Pumping}
\label{sec:App}

In this section we present two applications of the formalism developed above. As shown below, one of the interesting features of non-adiabatic pumping is an increasing delay in the current response to the external drive with growing driving speed. Hence, in distinction to the adiabatic limit, the current caused by a train of pulses can show interesting transient effects, whenever the pulse period is shorter than the system response time. To better understand non-adiabatic driving effects, we focus our analysis on single pulses and vary the speed by which their shape is changed.

It is worth stressing that our propagation method does not possess restrictions on the time dependence of the system driving parameters. In other words, the external time-dependent drive can be just a single pulse or a train of pulses, it can also be either fast or slow as compared with the system internal time scales.

\subsection{Symmetric monoparametric pumping}
\label{sec:comparison}

Let us begin by discussing the current generated by a single Gaussian voltage pulse changing the resonance energy as
\begin{align}
\label{eq:pumping1}
     \varepsilon_{\rm d}(t) & = \varepsilon_0 + \varepsilon_1 \exp\left[ -(t/\tpulse)^2 \right]\;.
\end{align}
Here $\tpulse$ sets the pumping time-scale. We take $\Gamma_\alpha(t)$ to be time-independent and equal for both leads, $\Gamma_{\rm L}(t)=\Gamma_{\rm R}(t)=\Gamma_0/2$.
Since thereby $J_{\rm L}=J_{\rm R}$, we will consider only $J_{\rm L}$ in the following.

\begin{figure}[t]
     \centering
     \includegraphics[width=.8\columnwidth]{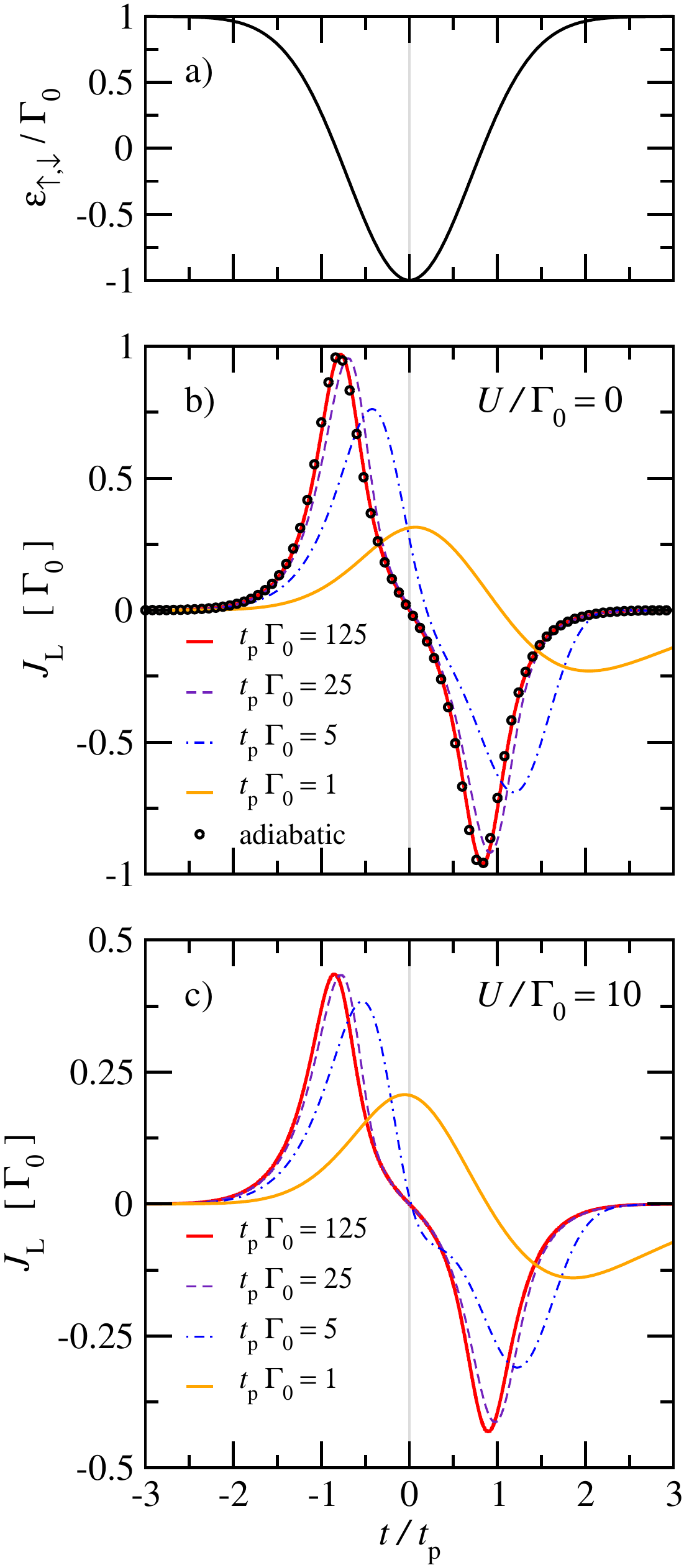}
     \caption{a) Time dependence of the resonance energy according to Eq.\ \eqref{eq:pumping1};
     b) and c) Time-resolved current $J_{\rm L}$ for 
     different pulse lengths $\tpulse$ and $U=0$ and $U/\Gamma_0=10$. 
	  Parameters used: $\varepsilon_0=\Gamma_0$, $\varepsilon_1=-2\Gamma_0$, $\mu_{\alpha} = 0$,
	   $k_{\rm B}T=0.1\Gamma_0$ and $N_\mathrm{F} = 160$ (number of auxiliary modes).
     Dots denote the adiabatic limit \cite{spgo+05}.
      }\label{fig:single}  
\end{figure}

Figure \ref{fig:single}a shows the time dependence of the resonance energy according to Eq.\ \eqref{eq:pumping1}. The two bottom panels show the instantaneous current $J_{\rm L}$ as a function of time for both the non-interacting ($U=0$) and the interacting ($U\neq 0$) case. In the limit of large 
$\tpulse$, we use as a check for our results an analytical expression for the pumped current $J_{\rm L}$, obtained for $U=0$ within the adiabatic approximation \cite{br98,spgo+05}.

Here, due to the L/R symmetry, there is no net charge flowing through the QD. At any given time both leads pump the same amount of charge in or out. In the driving scheme defined by Eq.\ \eqref{eq:pumping1}, the QD is initially nearly empty. At $t=0$, the resonance energy favors an almost full occupation. For very slow pumping, large $\tpulse \Gamma_0$, the current $J_{\rm L}$ depends only on the resonance energy $\varepsilon_{\rm d}(t)$: As the resonance dives into the Fermi sea, the QD is loaded with charge and the process is reversed as $\varepsilon_{\rm d}(t)$ starts increasing. This is no longer true when the drive is faster and $\tpulse\Gamma_0$ decreases: Now one observes a retardation effect, namely, the $J_{\rm L}$ depends not only on the resonance position, but also on driving speed. For fast driving one needs to integrate $J_{\rm L}$ over times much longer than $\tpulse$ to observe a vanishing net charge per pulse.

\subsection{Constrained two-parameter pumping}
\label{sec:monoparametric}
The pumped currents $J_{\rm L,R}(t)$ characterize the time-dependent electron response to the external drive. However, in most applications one is only interested in the charge pumped per cycle $Q_{\rm c}$ or per pulse $Q_{\rm p}$. In the latter case, $Q_{\rm p}$ is given as time integral over the current which we write in a symmetric way 
\begin{equation}\label{eq:chargepp}
     Q_{\rm p} = \frac{1}{2} \int\limits^{+\infty}_{-\infty} dt \left[ J_{\rm L}(t) -  J_{\rm R}(t) \right]\;.
\end{equation}
One of the beautiful lessons learned form the investigation of adiabatic pumping, establishes a proportionality relation between $Q_{\rm p}$ and the area swapped by the time-dependent driving forces in parameter space \cite{br98}. In other words, the total charge flowing through a QD per cycle (or per pulse) in a {\sl single-parameter adiabatic pump} vanishes.  
Due to the constraints of single-parameter pumps, in most applications at least two
parameters are used \cite{pola+92,mana+94,swma+99,dima+03,shso+06,fufa+07,buka+08,funi+08,blka+07,wrbl+08,kaka+08a,kaka+08,wrbl+09}.
On the other hand, by using a single-gate modulation one can realize
a constrained two-parameter pump \cite{kaka+08}, which implies that the time-dependence of the parameters is ultimately coupled due to the modulation of only a single gate voltage. In the following we will investigate the implications of this scenario for non-adiabatic pumping.

\begin{figure}[b]
\centering
\includegraphics[width=\columnwidth]{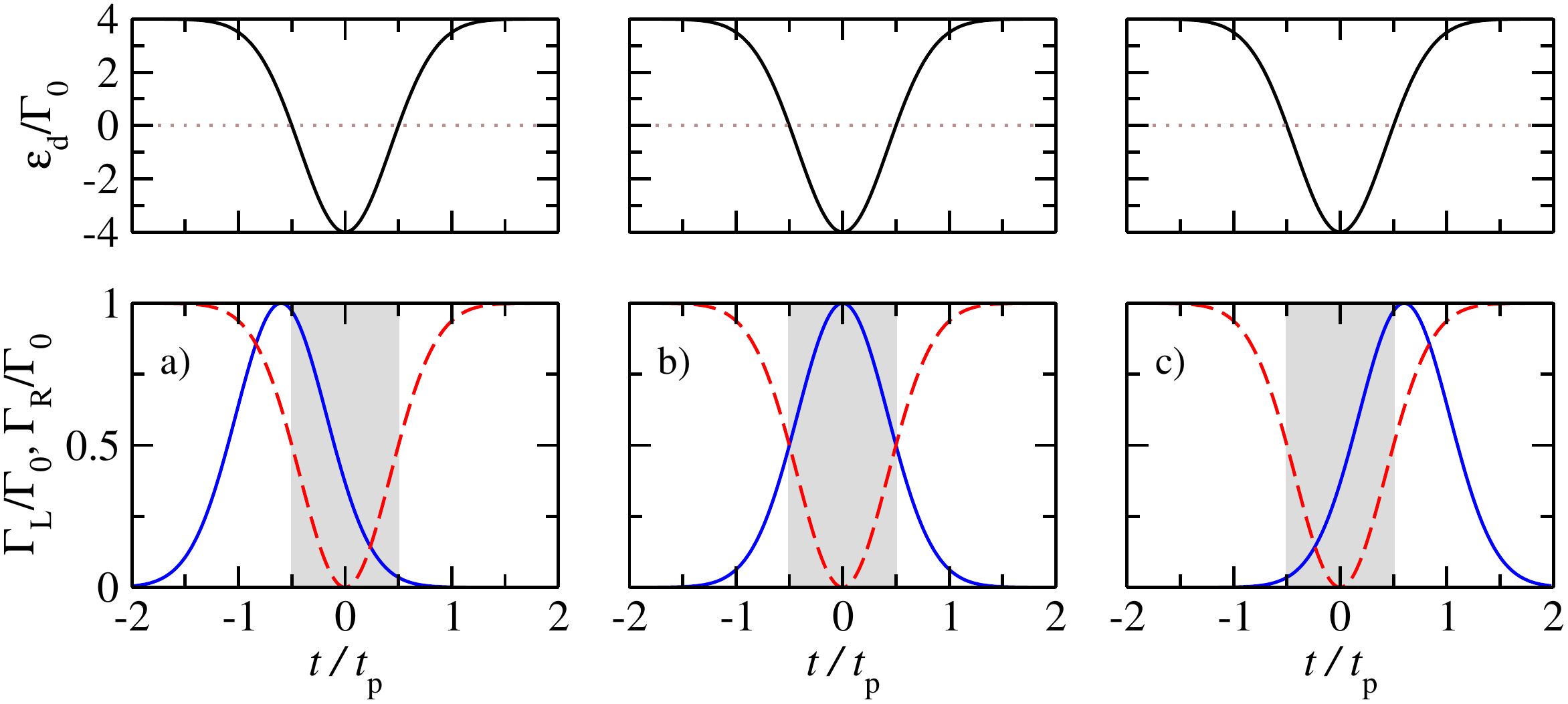}
\caption{Time-dependence of the resonance energy $\varepsilon_{\rm d}$ (upper row) and the decay widths $\Gamma_{\rm L}$ (lower row, blue/full line) and $\Gamma_{\rm R}$ (lower row, red/broken line) for three different cases: a) $\delta_{\rm L}=-1$, b) $\delta_{\rm L}=0$, and c) $\delta_{\rm L}=1$. The dotted lines indicate the chemical potential in the reservoirs and the shaded area shows the times when the resonance energy is below the chemical potential.
}
\label{fig:AppScheme3}  
\end{figure}%

\subsubsection{Pulse Scheme}
Specifically, let us consider voltage pulses of the form
\begin{equation}\label{eq:scheme3}
	S(t,\delta) = 1 - 2\exp\left[ -(2\sqrt{\ln2}\, t/\tpulse - \delta)^2  \right] \;.
\end{equation}
Here $\tpulse$ measures the characteristic pulse time, whereas $\delta$ governs the time the pulse sets in. 
The numerical factor $2\sqrt{\ln2}$ ensures that $\tpulse$ is the full width at half maximum of the pulse, which simplifies the following discussion.
By tuning the delay one can conveniently switch between a single parameter ($\delta=0$) and a two parameter setup ($\delta\neq0$).
Further, the time-dependence of the resonance energy and the coupling strengths (decay widths) are chosen as 
\begin{subequations}
\begin{align}
     \varepsilon_{\rm d}(t) & = \varepsilon_0 + \varepsilon_1 S(t, 0)\;,\\
     \Gamma_{\rm R}(t) & = \frac{\Gamma_0}{2} \left[ 1 -  S(t, 0) \right]\;,\\
     \Gamma_{\rm L}(t) & = \frac{\Gamma_0}{2} \left[ 1 +  S(t, \delta_{\rm L}) \right]\;.
\end{align}
\end{subequations}
This choice takes into account that the coupling strengths depend exponentially on
the gate voltage \cite{kaka+08}. The constraint is imposed by setting $\delta_{\rm R}=0$ and the specific value of $\delta_{\rm L}$.
For this driving parameterization, the resonance and the decay widths are $\varepsilon_0$ and $\Gamma_0$, respectively, for both asymptotic limits of $|t| \gg \tpulse$. In the following, the parameters are taken as $\varepsilon_0 = 0$, $\varepsilon_1/\Gamma_0 = 4$, $k_{\rm B}T/\Gamma_0 = 1/10$, and interaction energy either $U=0$ or $U/\Gamma_0=10$. 
In Fig.\ \ref{fig:AppScheme3} the time dependence of $\varepsilon_{\rm d}$ and $\Gamma_{\rm L/R}$ is illustrated for three cases $\delta_{\rm L}=0$ and $\delta_{\rm L}=\pm1$. As mentioned above, in each case the coupling to the right reservoir $\Gamma_{\rm R}$ follows the time dependence of the resonance energy. When the latter attains its minimal value at $t=0$, which brings the energy well below the chemical potential of the reservoirs, the coupling to the right reservoir is {\it minimal}. On the other hand, the behavior of the coupling to the left reservoir can be influenced by the value of $\delta_{\rm L}$.
For $\delta_{\rm L}=-1$ the {\it maximum} of $\Gamma_{\rm L}$ comes before $t=0$, while for 
$\delta_{\rm L}=+1$ it is attained after $t=0$. In the case $\delta_{\rm L}=0$ the coupling to the left
reservoir is maximal simultaneously with $\Gamma_{\rm R}$ being minimal at $t=0$. In the following the response to these drivings will be investigated.

\subsubsection{Adiabatic Pumping}
\begin{figure}[b]
\centering
\includegraphics[width=.8\columnwidth]{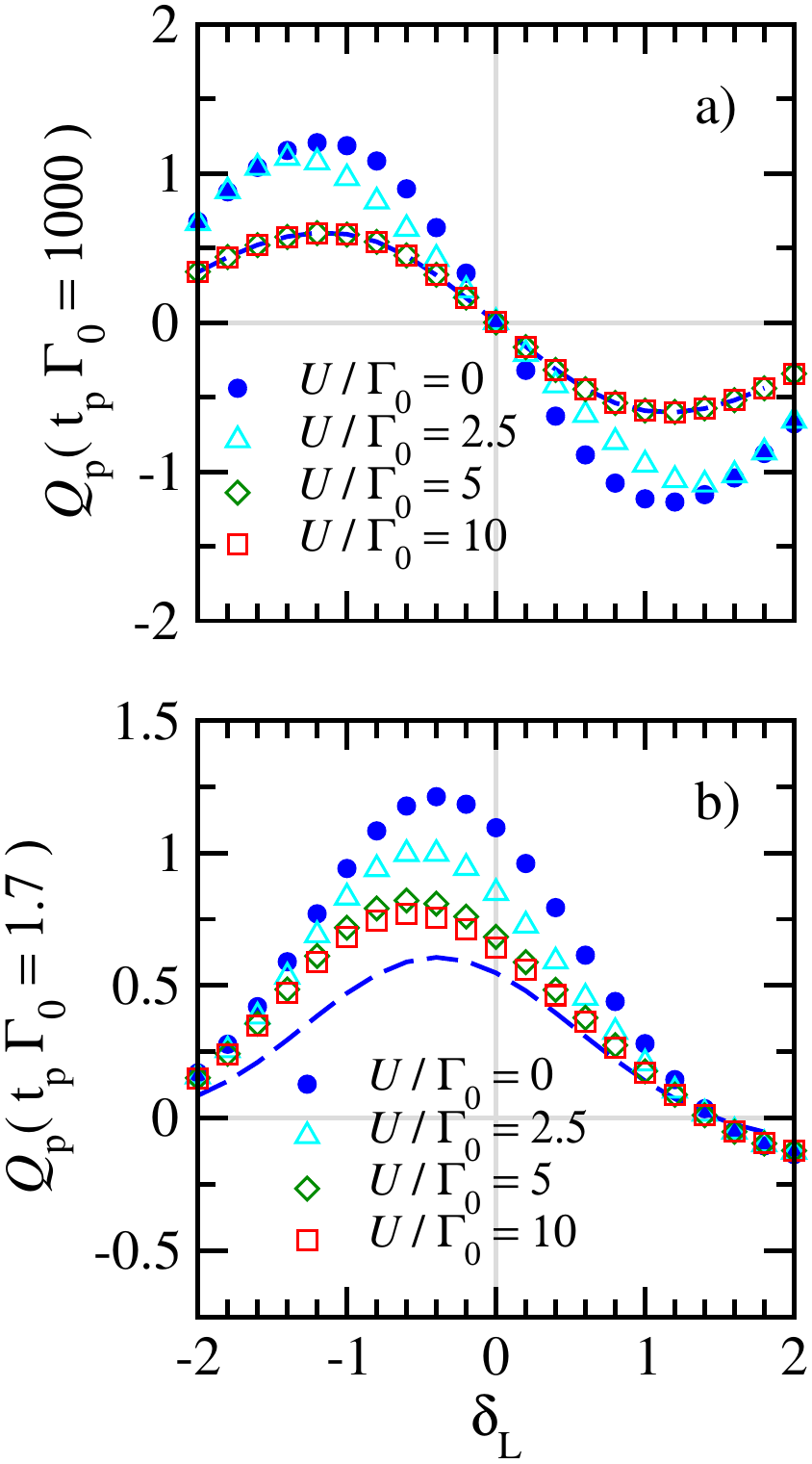}
\caption{Charge per pulse $Q$ vs pulse-shift $\delta_{\rm L}$ in
	the long-pulse limit (upper panel) and at $\tpulse = 1.7\,\Gamma^{-1}_0$ (lower panel). 
	The dashed lines indicate half of the non-interacting result.
}
\label{fig:AppMonoQpPhi}  
\end{figure}%
\begin{figure*}[t]
\centering
\includegraphics[width=\textwidth]{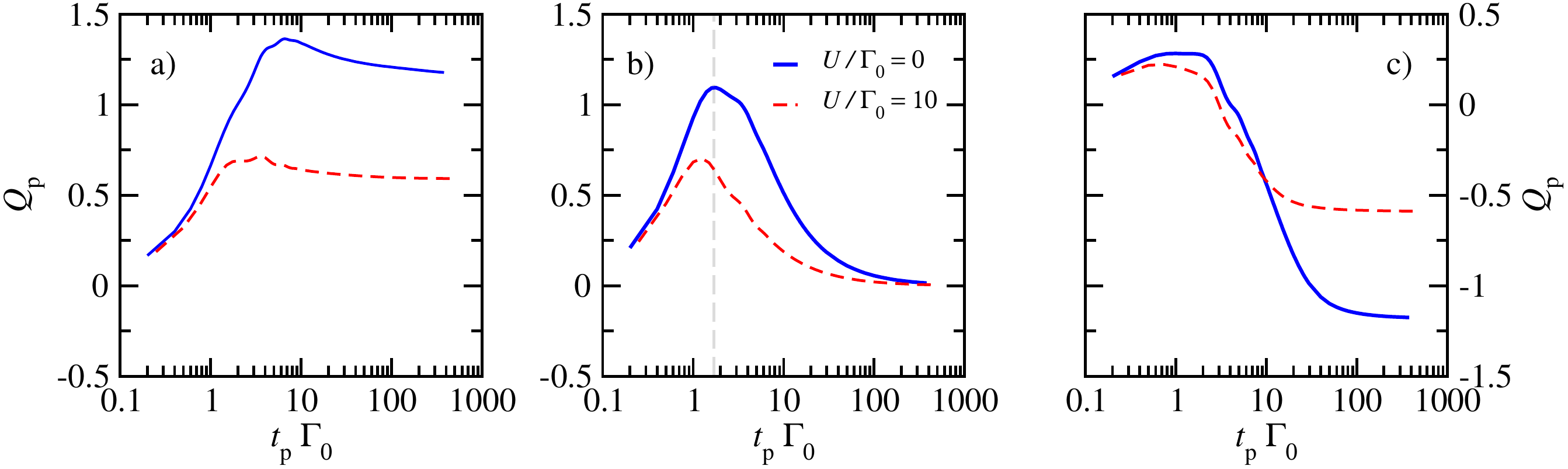}
\caption{Charge pumped per pulse $Q_{\rm p}$ versus pulse-length $\tpulse$ for the pulse scheme given by Eq.\ \eqref{eq:scheme3}. The (blue) solid line stands for the non-interacting case, while the (red) broken line represents $U/\Gamma_0=10$. We consider three different cases: a) $\delta_{\rm L}=-1$, b) $\delta_{\rm L}=0$, and c) $\delta_{\rm L}=1$. 
}
\label{fig:AppMonoQp}  
\end{figure*}%

Knowing the time dependence shown in Fig.\ \ref{fig:AppScheme3} one can readily predict the behavior of $Q_{\rm p}$ in the adiabatic limit. 
In this case, electron flow occurs when the resonance energy matches the chemical potential of the reservoirs. In our pulse scheme, $\varepsilon_{\rm d}(t)$ equals the chemical potential at $t=-\tpulse/2$ and $t=+\tpulse/2$ corresponding to the onset of charging and de-charging of the QD, respectively. Further, the direction of the net current is determined by the difference of the
couplings to the reservoirs at these very times. For example, for $\delta_{\rm L}=-1$ one finds
$\Gamma_{\rm L}>\Gamma_{\rm R}$ while charging and $\Gamma_{\rm L}<\Gamma_{\rm R}$ while de-charging. Consequently, the net current is directed from left to right and $Q_{\rm p}$ is
expected to be {\it positive}. For $\delta_{\rm L}=+1$ the situation is opposite and 
$Q_{\rm p}$ should be {\it negative}. Finally, for $\delta_{\rm L}=0$ the couplings are equal
at both instants of time and the net current is vanishing. 
These expectations are confirmed by our results for the adiabatic regime, 
$\tpulse\Gamma_0\gg 1$, and different values of $U/\Gamma_0$, which are shown in Fig.\ 
\ref{fig:AppMonoQpPhi}a. As already mentioned, one observes $Q_{\rm p} =0$ for $\delta_{\rm L}=0$ (monoparametric pumping). As $|\delta_{\rm L}|$ begins to increase, $|Q_{\rm p}|$ increases as well. In this scenario, when the resonance energy matches the chemical potential, electrons load the dot from the left (or right) and later they are unloaded to the right (or left, depending on the sign on $\delta$). For larger values of $|\delta_{\rm L}|$, the left reservoir participates less in the loading or unloading of the QD and the charge per pulse vanishes accordingly.

For interaction strengths $U\gg\varepsilon_1$ the double occupation of the QD is suppressed
and consequently, in the adiabatic regime, $Q_{\rm p}$ is half the value of $Q_{\rm p}$ for the non-interacting case. The numerical results indicate that within the Hubbard I approximation, $U\neq 0$ does not introduce new time scales to the problem for $\tpulse\Gamma_0\gg 1$, and its major effect is to correct the spin degeneracy factor in the equations for the $U=0$ case.

\subsubsection{Non-adiabatic Pumping}
None of the aforementioned features are observed in the non-adiabatic pumping regime. Figure \ref{fig:AppMonoQpPhi}b shows, for example, that, for short pulses there is no simple relation between $Q_{\rm p}$ for $U=0$ and for $U\neq 0$. Moreover, compared to the adiabatic regime
the charge per pulse can be substantially larger in this regime. Unfortunately, the behavior of
$Q_{\rm p}$ in this regime is not as easily predicted in general, since the evolution of the
parameters $\{ \varepsilon_{\rm d}$, $\Gamma_{\rm L}$, $\Gamma_{\rm R}\}$ after the onset of
loading and unloading has to be taken into account. This is because in the non-adiabatic regime the QD charging and de-charging is delayed with respect to the external system changes, as it was shown in Sec.\ \ref{sec:comparison}.
Taking, for example, the case $\delta_{\rm L}=0$ one finds from Fig.\ \ref{fig:AppScheme3}b, that $\Gamma_{\rm L}>\Gamma_{\rm R}$ while the resonance energy is below the chemical potential and charging occurs. During the de-charging, when $\varepsilon_{\rm d}>\mu_{\rm L,R}$, one finds $\Gamma_{\rm R}>\Gamma_{\rm L}$. Consequently, the current is expected to flow
mainly from left to right, which leads to a positive charge per cycle. This is confirmed
by the results shown in Fig.\ \ref{fig:AppMonoQpPhi}b. The quantitative behavior
depends on the precise magnitude of the delay, which is determined by the pulse length. However, from the analysis presented above and for sufficiently short pulses one concludes that $Q_{\rm p}$ has to be positive independent of $\delta_{\rm L}$. The interesting implications of
this result will be discussed at the end of this section.
 
Finally, in Fig.\ \ref{fig:AppMonoQp} we summarize and corroborate the discussion of the
non-adiabatic pumping. It shows the charge pumped due the pulse as a function of pulse length $\tpulse\Gamma_0$ in the non-interacting ($U=0$) and the Coulomb blockade regime ($U=10\Gamma_0$). In the latter case $Q_{\rm p} \le 1$ for all pulse lengths. 
As discussed above, the amount of pumped charge $Q_{\rm p}$ depends very strongly on the value of $\tpulse\Gamma_0$. In the limit of large pulse lengths, $Q_{\rm p}$ approaches the
respective adiabatic value, while for $\tpulse\Gamma_0\to0$ the charge per pulse vanishes.
Moreover, one finds that $Q_{\rm p}$ is indeed positive for small pulse lengths. This
has the intriguing consequence that the charge per pulse can change its sign sweeping
from short to long pulses. This is shown in Fig.\ \ref{fig:AppMonoQp}b for $\delta_{\rm L}=1$,
where $Q_{\rm p}$ is negative in the adiabatic regime. A more general and quantitative analysis
of this effect is certainly desirable, but beyond the scope of this article. It may lead, however, to interesting new applications.
It is also worth to mention, that by changing the pumping parameters it is possible to optimize the charge pumped per pulse and in particular to find situations where $Q_{\rm p}=1$, which may be very interesting for metrology purposes \cite{ke08}.

\section{Conclusions}
\label{sec:conclusion}
We presented a new method to analyze non-adiabatic charge pumping through single-level quantum dots that takes into account Coulomb interactions. The method is based on calculating the time evolution for single-electron density matrices. The many-body aspects of the problem are approximated by truncating the equations of motion one order beyond mean field. The novelty is the way the time evolution is treated: By means of an auxiliary-mode expansion, we obtain a propagation scheme that allows for dealing with arbitrary driving parameters, fast and slow. 
The method presented in this paper can be applied to a wide range of coupling parameters $\Gamma_\alpha$, provided one avoids the Kondo regime. Hence, we are not restricted to the weak coupling limit where $Q_{\rm p}$, the charge pumped per pulse, is rather small.

The presented results for single-pulses are also valid for pulse trains, provided the time between the pulses is sufficiently long. One can expect to find 
qualitatively new and interesting effects by decreasing the time lag. The propagation scheme  allows, in principle, for studying transient effects. In addition, by propagating over a periodic sequence of pulses it constitutes a complementary approach to the more familiar periodic driving. In this regard, our propagation scheme has the potential to be a valuable tool and provide deeper insights into non-adiabatic quantum pumps.

\begin{acknowledgments}
This work is supported in part by CNPq (Brazil).
\end{acknowledgments}
  
\appendix
\section{Auxiliary-mode expansion}
\label{sec:AppExp}

Here we motivate the rules given in Sect.\ \ref{sec:Prop}.
To begin with we introduce correlation functions, which can be approximated by finite sums. 
Then we write the current matrices in terms of these finite sums. 
\subsection{Correlation functions and mode expansion}
As we will show later, in the present case we have to consider the following 
reservoir correlation function
\begin{align}
	C_\alpha (t',t) 
		\equiv {} & \sum_k T_{\alpha k} (t) g_{\alpha k}(t',t) T^*_{\alpha k}(t') \fermi_{\alpha}(\varepsilon_{k})\notag\\
		= {} & \int \frac{d\varepsilon}{2 \pi}\,
		\Gamma_{\alpha}(\varepsilon,t',t)\, \fermi_{\alpha}(\varepsilon)\,\notag\\
		 &\times
      \exp\left\{
      		\imath \int\limits_{t'}^{t} dt'' [\varepsilon+\Delta\varepsilon_\alpha (t'')]
		\right\} \;,\label{eq:CorrFuncInt}
\end{align}
where the line-width function $\Gamma_{\alpha}$ is defined as usual \cite{haja07}
\begin{align}
\Gamma_{\alpha}(\varepsilon,t',t) & \equiv 2\pi \sum_k T_{\alpha k} (t) T^*_{\alpha k}(t')
												\delta(\varepsilon - \varepsilon_{\alpha k})
			\notag\\ 
			& = 2\pi T_{\alpha} (t) T^*_{\alpha}(t') \rho_\alpha.
\end{align}
In the second line we have used the wide-band limit.

In order to perform the energy integration in Eq.~\eqref{eq:CorrFuncInt} we
expand the Fermi function $\fermi(\varepsilon)$
as a finite sum over simple poles
\begin{equation}
 \fermi(\varepsilon)
  \approx \frac{1}{2} - \frac{1}{\beta}\sum_{p=1}^{N_\mathrm{F}}\left(
    \frac{1}{\varepsilon{-}\chi_{p}^+}
    +\frac{1}{\varepsilon{-}\chi_{p}^-}\right)\,,
  \label{eq:ExpFermiFun}
\end{equation}
with $\chi_{p}^\pm = \mu{\pm}x_p/\beta$ and $\Im x_p >0$. Instead of using the Matsubara expansion \cite{ma90}, with poles $x_p=\imath\pi(2p{-}1)$, we use a partial fraction decomposition of the 
Fermi function \cite{crsa09crsa10}, which converges much faster than the standard Matsubara expansion. In this case the poles $x_p=\pm 2\sqrt{z_p}$ are given by the eigenvalues $z_p$ of a ${N_\mathrm{F}}{\times}{N_\mathrm{F}}$ matrix \cite{crsa09crsa10}.
The poles are arranged such that all poles $\chi_p^+$ ($\chi_p^-$) are in the upper (lower) complex plane.
As in the Matsubara expansion all poles have the same weight.

Employing the expansion given by Eq.\ \eqref{eq:ExpFermiFun}, one can evaluate the energy integrals
by contour integration in the upper or lower complex plane depending on the sign of $t-t'$. 
Thereby, the integral in Eq.~\eqref{eq:CorrFuncInt} becomes a (finite) sum of the residues. For $t\ge t'$ one gets
\begin{subequations}\label{eq:CorrFuncExp}
\begin{align}
  C_\alpha (t',t) 
  = {} &  \frac{1}{2} \Gamma_\alpha(t) \delta(t-t') 
  \nonumber\\ & +\, T_\alpha(t)\, \sum\limits_p C_{\alpha p}(t,t_{1})\;,\\
  C_{\alpha p}(t',t) \equiv {} &
  \frac{-\imath}{\beta}  T_\alpha(t')
      e^{\imath \int^t_{t'} dt'' \chi^+_{\alpha p} (t'')}\;,
      \label{eq:PartSelfEnerg}
\end{align}
\end{subequations}
with the auxiliary modes for reservoir $\alpha$ given by
\begin{equation}\label{eq:auxmodesdef}
	\chi^+_{\alpha p} (t) = [\mu_{\alpha}+ \Delta\varepsilon_\alpha (t)]+x_p/\beta.
\end{equation}
Here, $\mu_{\alpha}$ is the chemical potential and $\Delta\varepsilon_\alpha (t)$ 
is due to the time-dependent single-particle energies 
$\varepsilon_{\alpha k}(t)$ of the reservoir Hamiltonian [Eq.\,(\ref{eq:ResHamilOp})].

\subsection{Current matrices}
The set of equations \eqref{eq:Pi2EOM} and \eqref{eq:PiEOMHubbard} can be formally solved.
In order to write down these solutions we define the following functions
\begin{subequations}
\def\exp#1{\mathrm{e}^{#1}}
\begin{align}
	g_s (t,t') & \equiv \exp{
	- \imath \int^t_{t'} dt'' \left[\varepsilon_s(t'') 
	- \imath \frac{\Gamma(t'')}{2}\right]
	} \;,\\
	g^U_s (t,t') & \equiv \exp{
	- \imath \int^t_{t'} dt'' \left[\varepsilon_s(t'') + U
	- \imath \frac{\Gamma(t'')}{2}\right]
	}\;.
\end{align}
\end{subequations}
With these definitions the formal solution of Eq.\ \eqref{eq:Pi2EOM} reads
\begin{align}
\PIss_{\alpha k s}(t) = {} & \int^t_{t_0} dt' g_s(t,t') g_{\alpha k}(t',t)
\times\notag\\ &\times
 \left[
	T^*_{\alpha k}(t')\fermi_{\alpha}(\varepsilon_{k}) 
 -\imath U \PFHI_{\alpha k s}(t') \right]\,,
\end{align}
where we have assumed $\PIss_{\alpha k s}(t_{0})=0$, corresponding to our choice of an initially uncorrelated density matrix (see Sec.\ \ref{sec:setup}).
An analogous equation holds for $\PFHI_{\alpha k s}(t)$, again with $\PFHI_{\alpha k s}(t_{0})=0$.
We can combine these two expressions to get for the second part of the current matrix
\begin{widetext}
\begin{align}
\PIss_{\alpha s}(t) = \sum_k T_{\alpha k}(t) \PIss_{\alpha k s}(t) 
 & =	\int\limits^t_{t_0} dt' C_\alpha (t',t) g_s(t,t') 
-\imath U \int\limits^t_{t_0} dt' g_s(t,t') \int\limits^{t'}_{t_0} dt''
	 C_\alpha (t'',t) g^U_s(t',t'') n_{\bar{s}}(t'') \;,
\end{align}
where we have used the definition of the correlation function $C_\alpha$ given by Eq.\ \eqref{eq:CorrFuncInt}.
Finally, by means of the expansion \eqref{eq:CorrFuncExp} of the correlation functions we obtain an expansion of the current matrices
\begin{subequations}
\begin{align}
\sum_k T_{\alpha k}(t) \PIss_{\alpha k s}(t) 
 & =  \frac{1}{4} \Gamma_\alpha(t) + T_\alpha(t)\, \sum\limits_p \PIss_{\alpha s p}(t)\;,
\\
\PIss_{\alpha s p}(t) &\equiv \int\limits^t_{t_0} dt' C_{\alpha p} (t',t) g_s(t,t') 
-\imath U \int\limits^t_{t_0} dt' g_s(t,t') \int\limits^{t'}_{t_0} dt''
	 C_{\alpha p} (t'',t) g^U_s(t',t'') n_{\bar{s}}(t'')	\;,
\end{align}
\end{subequations}
\end{widetext}
which resembles the last rule of Eqs.\ \eqref{eq:rules}.
Using the explicit expression for $\PIss_{\alpha s p}$ and taking the time derivative
one can easily verify the first two rules given by Eqs.\ \eqref{eq:rules}. Similarly, one also obtains an expression for $\PFHI_{\alpha s p}$, which reads
\begin{equation}
\PFHI_{\alpha s p} (t) \equiv
\int\limits^{t}_{t_0} dt''
	 C_{\alpha p} (t'',t) g^U_s(t,t'') n_{\bar{s}}(t'')\;.
\end{equation}
The time derivative of this expression is given by Eq.\ \eqref{eq:AuxEOMTPhi}.

\section{Stationary occupations}
\label{sec:AppStat}
If neither the couplings $T_{\alpha k}$ (and thus $\Gamma$) nor the levels $\varepsilon_{s}$ or $\varepsilon_{\alpha k}$ depend on time the level occupations $n_{s}$ and the currents $J_{\alpha}$ converge to stationary values. These values can be obtained by setting all time derivatives in the respective equations of motion to zero.
In order to simplify the notation we characterize the stationary values by omitting the time argument.

We will specify below the level occupations for the two approximations discussed in Sec.\ \ref{sec:ExpVal}.
Therefore we use
\begin{equation}\label{eq:OccAndPIss}
	n_{s} = \frac{1}{\Gamma} 2\Re \sum_{\alpha k}  T_{\alpha k} \PIss_{\alpha k s} \,,
\end{equation}
which follows directly from Eq.\ (\ref{eq:NEGFDefPi1}b).

\subsection{Hartree-Fock}
Within the Hartree-Fock approximation [Sec.\ \ref{sec:HFapp}] we get from Eq. \eqref{eq:PiEOMHartree}
\begin{equation}
	\PIss_{\alpha k s} =
	\imath\frac{T^*_{\alpha k} f_{\alpha k}}{
	\Eaks - U n_{\bar{s}}}\,.
\end{equation}
Plugging this into Eq.\ \eqref{eq:OccAndPIss}, changing the $k$ summation into an integral over $\varepsilon$ and using the definition \eqref{eq:Deltaks}, we get for the wide-band limit [Eq.\ \eqref{eq:GammaWBL}] 
\begin{align}\label{eq:StationaryHF}
n_{s} 
	=& \sum_\alpha \Gamma_\alpha \int \frac{d\varepsilon}{2\pi}
		\frac{\fermi_{\alpha}(\varepsilon)}{
		\left(\varepsilon{-}\varepsilon_s{-} U\,n_{\bar{s}} \right)^2
		+ \left(\frac{\Gamma}{2}\right)^2
		}\,.
\end{align}
Equation  \eqref{eq:StationaryHF} is a non-linear equation for $n_{s}$ and has to be solved numerically.

\subsection{Hubbard I}
We obtain the stationary conditional current $\PFHI_{\alpha k s}$ for the Hubbard I approximation [Sec.\ \ref{sec:HIapp}] from Eq.\ \eqref{eq:PiEOMHubbard} as
\begin{equation}
\PFHI_{\alpha k s} =
	\imath\frac{T^*_{\alpha k} f_{\alpha k} n_{\bar{s}}}{
	\Eaks -U +\imath\Gamma
		}\,.
\end{equation}
This expression can be used for the stationary $\PIss$
in Eq.\ \eqref{eq:Pi2EOM}
\begin{align}
\PIss_{\alpha k s}
		= {} & \imath\, \frac{T^*_{\alpha k}\, f_{\alpha k}}{\Eaks}
		 +\imath \frac{T^*_{\alpha k}\, f_{\alpha k}\; U n_{\bar{s}}
	}{
	\Eaks\left[\Eaks-U+\imath\Gamma\right]
	}
\end{align}
We use Eq.\ \eqref{eq:OccAndPIss} and the definition \eqref{eq:Deltaks} and finally get for the occupation the following integral
\begin{subequations}
\begin{align}\label{eq:StationaryHI}
n_{s} 
	&= \sum_\alpha \Gamma_\alpha \int \frac{d\varepsilon}{2\pi}\fermi_{\alpha}(\varepsilon)
	 \left[A'(\varepsilon)+n_{\bar{s}}A''(\varepsilon)\right]\\
	A'(\varepsilon) &\equiv 
	\frac{1}{(\varepsilon{-}\varepsilon_{s})^{2}+\left(\frac{\Gamma}{2}\right)^{2}}
	\\
		A''(\varepsilon) &\equiv A'(\varepsilon)
		\frac{U\left[4(\varepsilon{-}\varepsilon_{s})-U\right]}{
		(\varepsilon{-}\varepsilon_{s}{-}U)^{2}+\left(\frac{3\Gamma}{2}\right)^{2}
		}\,.
\end{align}
\end{subequations}
This time the equation is linear in $n_{\bar{s}}$ and can be solved explicitly.
In the limits $U\to0$ and $U\to\infty$ it is $A''(\varepsilon)=0$ and $A''(\varepsilon)=-A'(\varepsilon)$, respectively. The former limit corresponds to non-interacting electrons
and Eq.\ \eqref{eq:StationaryHI} gives the correct expression for the occupation \cite{haja07}.
The latter case describes the situation with very strong interactions.



\begin{thebibliography}{10}

\bibitem{th83}
D.~J. Thouless, Phys. Rev. B {\bf 27},  6083  (1983).

\bibitem{pola+92}
H. Pothier, P. Lafarge, C. Urbina, D. Esteve, and M. Devoret, Europhys. Lett.
  {\bf 17},  249  (1992).

\bibitem{mana+94}
J.~M. Martinis, M. Nahum, and H.~D. Jensen, Phys. Rev. Lett. {\bf 72},  904
  (1994).

\bibitem{swma+99}
M. Switkes, C.~M. Marcus, K. Campman, and A.~C. Gossard, Science {\bf 283},
  1905  (1999).

\bibitem{dima+03}
L. DiCarlo, C. Marcus, and J. J.~S.~Harris, Phys. Rev. Lett. {\bf 91},  246804
  (2003).

\bibitem{lebu+05}
P.~J. Leek, M.~R. Buitelaar, V.~I. Talyanskii, C.~G. Smith, D. Anderson,
  G.~A.~C. Jones, J. Wei, and D.~H. Cobden, Phys. Rev. Lett. {\bf 95},  256802
  (2005).

\bibitem{wapo+03}
S.~K. Watson, R.~M. Potok, C.~M. Marcus, and V. Umansky, Phys. Rev. Lett. {\bf
  91},  258301  (2003).

\bibitem{shso+06}
Y.-S. Shin, W. Song, J. Kim, B.-C. Woo, N. Kim, M.-H. Jung, S.-H. Park, J.-G.
  Kim, K.-H. Ahn, and K. Hong, Phys. Rev. B {\bf 74},  195415  (2006).

\bibitem{fufa+07}
A. Fuhrer, C. Fasth, and L. Samuelson, Appl. Phys. Lett. {\bf 91},  052109
  (2007).

\bibitem{buka+08}
M.~R. Buitelaar, V. Kashcheyevs, P.~J. Leek, V.~I. Talyanskii, C.~G. Smith, D.
  Anderson, G.~A.~C. Jones, J. Wei, and D.~H. Cobden, Phys. Rev. Lett. {\bf
  101},  126803  (2008).

\bibitem{funi+08}
A. Fujiwara, K. Nishiguchi, and Y. Ono, Appl. Phys. Lett. {\bf 92},  042102
  (2008).

\bibitem{br98}
P.~W. Brouwer, Phys. Rev. B {\bf 58},  R\,10135  (1998).

\bibitem{zhsp+99}
F. Zhou, B. Spivak, and B. Altshuler, Phys. Rev. Lett. {\bf 82},  608  (1999).

\bibitem{enah+02}
O. Entin-Wohlman, A. Aharony, and Y. Levinson, Phys. Rev. B {\bf 65},  195411
  (2002).

\bibitem{albr+02}
I.~L. Aleiner, P.~W. Brouwer, and L.~I. Glazman, Phys. Rep. {\bf 358},  309
  (2002).

\bibitem{mobu01}
M. Moskalets and M. B\"uttiker, Phys. Rev. B {\bf 64},  201305(R)  (2001).

\bibitem{mami01}
Y. Makhlin and A.~D. Mirlin, Phys. Rev. Lett. {\bf 87},  276803  (2001).

\bibitem{roci10}
F. Romeo and R. Citro, Phys. Rev. B {\bf 82},  085317  (2010).

\bibitem{shch01}
P. Sharma and C. Chamon, Phys. Rev. Lett. {\bf 87},  096401  (2001).

\bibitem{much+02}
E.~R. Mucciolo, C. Chamon, and C.~M. Marcus, Phys. Rev. Lett. {\bf 89},  146802
   (2002).

\bibitem{male+04}
M. Mart\'{\i}nez-Mares, C.~H. Lewenkopf, and E.~R. Mucciolo, Phys. Rev. B {\bf
  69},  085301  (2004).

\bibitem{vaam+01}
M.~G. Vavilov, V. Ambegaokar, and I.~L. Aleiner, Phys. Rev. B {\bf 63},  195313
   (2001).

\bibitem{mobu02}
M. Moskalets and M. B{\"u}ttiker, Phys. Rev. B {\bf 66},  205320  (2002).

\bibitem{armo06}
L. Arrachea and M. Moskalets, Phys. Rev. B {\bf 74},  245322  (2006).

\bibitem{mule07}
E.~R. Mucciolo and C.~H. Lewenkopf, Int. J. Nanotechnol. {\bf 4},  482  (2007).

\bibitem{cian+03}
R. Citro, N. Andrei, and Q. Niu, Phys. Rev. B {\bf 68},  165312  (2003).

\bibitem{ao04}
T. Aono, Phys. Rev. Lett. {\bf 93},  116601  (2004).

\bibitem{spgo+05}
J. Splettstoesser, M. Governale, J. K{\"o}nig, and R. Fazio, Phys. Rev. Lett.
  {\bf 95},  246803  (2005).

\bibitem{seor06}
E. Sela and Y. Oreg, Phys. Rev. Lett. {\bf 96},  166802  (2006).

\bibitem{fisi08}
D. Fioretto and A. Silva, Phys. Rev. Lett. {\bf 100},  236803  (2008).

\bibitem{arye+08}
L. Arrachea, A.~L. Yeyati, and A. Martin-Rodero, Phys. Rev. B {\bf 77},  165326
   (2008).

\bibitem{hepi+09}
A.~R. Hern\'{a}ndez, F.~A. Pinheiro, C.~H. Lewenkopf, and E.~R. Mucciolo, Phys.
  Rev. B {\bf 80},  115311  (2009).

\bibitem{hawe+01}
B.~L. Hazelzet, M.~R. Wegewijs, T.~H. Stoof, and Y.~V. Nazarov, Phys. Rev. B
  {\bf 63},  165313  (2001).

\bibitem{saco+06}
R. S{\'a}nchez, E. Cota, R. Aguado, and G. Platero, Phys. Rev. B {\bf 74},
  035326  (2006).

\bibitem{brbu08}
M. Braun and G. Burkard, Phys. Rev. Lett. {\bf 101},  036802  (2008).

\bibitem{blha+95}
R.~H. Blick, R.~J. Haug, D.~W. van~der Weide, K. von Klitzing, and K. Eberl,
  Appl. Phys. Lett. {\bf 67},  3924  (1995).

\bibitem{ooko+97}
T.~H. Oosterkamp, L.~P. Kouwenhoven, A.~E.~A. Koolen, N.~C. van~der Vaart, and
  C.~J. P.~M. Harmans, Phys. Rev. Lett. {\bf 78},  1536  (1997).

\bibitem{to05}
L.~E.~F. Torres, Phys. Rev. B {\bf 72},  245339  (2005).

\bibitem{cago+09}
F. Cavaliere, M. Governale, and J. K{\"o}nig, Phys. Rev. Lett. {\bf 103},
  136801  (2009).

\bibitem{blka+07}
M.~D. Blumenthal, B. Kaestner, L. Li, S. Giblin, T.~J. B.~M. Janssen, M.
  Pepper, D. Anderson, G. Jones, and D.~A. Ritchie, Nature Phys. {\bf 3},  343
  (2007).

\bibitem{wrbl+08}
S.~J. Wright, M.~D. Blumenthal, G. Gumbs, A.~L. Thorn, M. Pepper, T.~J. B.~M.
  Janssen, S.~N. Holmes, D. Anderson, G.~A.~C. Jones, C.~A. Nicoll, and D.~A.
  Ritchie, Phys. Rev. B {\bf 78},  233311  (2008).

\bibitem{kaka+08a}
B. Kaestner, V. Kashcheyevs, G. Hein, K. Pierz, U. Siegner, and H.~W.
  Schumacher, Appl. Phys. Lett. {\bf 92},  192106  (2008).

\bibitem{kaka+08}
B. Kaestner, V. Kashcheyevs, S. Amakawa, M.~D. Blumenthal, L. Li, T.~J. B.~M.
  Janssen, G. Hein, K. Pierz, T. Weimann, U. Siegner, and H.~W. Schumacher,
  Phys. Rev. B {\bf 77},  153301  (2008).

\bibitem{wrbl+09}
S.~J. Wright, M.~D. Blumenthal, M. Pepper, D. Anderson, G.~A.~C. Jones, C.~A.
  Nicoll, and D.~A. Ritchie, Phys. Rev. B {\bf 80},  113303  (2009).

\bibitem{zhma+05}
Y. Zhu, J. Maciejko, T. Ji, H. Guo, and J. Wang, Phys. Rev. B {\bf 71},  075317
   (2005).

\bibitem{kust+05}
S. Kurth, G. Stefanucci, C.-O. Almbladh, A. Rubio, and E.~K.~U. Gross, Phys.
  Rev. B {\bf 72},  035308  (2005).

\bibitem{wesc+06}
S. Welack, M. Schreiber, and U. Kleinekath\"ofer, J. Chem. Phys. {\bf 124},
  044712  (2006).

\bibitem{mogu+07}
V. Moldoveanu, V. Gudmundsson, and A. Manolescu, Phys. Rev. B {\bf 76},  085330
   (2007).

\bibitem{myst+09}
P. My\"{o}h\"{a}nen, A. Stan, G. Stefanucci, and R. van Leeuwen, Phys. Rev. B
  {\bf 80},  115107  (2009).

\bibitem{crsa09a}
A. Croy and U. Saalmann, Phys. Rev. B {\bf 80},  245311  (2009).

\bibitem{meta99}
C. Meier and D.~J. Tannor, J. Chem. Phys. {\bf 111},  3365  (1999).

\bibitem{jizh+08}
J. Jin, X. Zheng, and Y. Yan, J. Chem. Phys. {\bf 128},  234703  (2008).

\bibitem{crsa11}
A. Croy and U. Saalmann, New J. Phys. {\bf 13},  043015  (2011).

\bibitem{crsa09crsa10}
A. Croy and U. Saalmann, Phys. Rev. B {\bf 80},  073102  (2009); Phys. Rev. B {\bf 82},  159904(E)  (2010).

\bibitem{jawi+94}
A.-P. Jauho, N.~S. Wingreen, and Y. Meir, Phys. Rev. B {\bf 50},  5528  (1994).

\bibitem{caco+71}
C. Caroli, R. Combescot, P. Nozieres, and D. Saint-James, J. Phys. C {\bf 4},
  916  (1971).

\bibitem{haja07}
H. Haug and A.-P. Jauho, {\em Quantum Kinetics in Transport and Optics of
  Semiconductors} (Springer, Berlin, 2007).

\bibitem{ke08}
M.~W. Keller, Metrologia {\bf 45},  102  (2008).

\bibitem{ma90}
G.~D. Mahan, {\em Many Particle Physics}, 2nd ed. (Plenum, New York, 1990).

\end{thebibliography}
\end{document}